\documentclass[aps,prb,twocolumn,showpacs,superscriptaddress,longbibliography]{revtex4-1}
\usepackage[plainpages=false,pdfpagelabels,colorlinks=true,linkcolor=red,urlcolor=blue,citecolor=blue,pdftitle={Title},pdfauthor={},pdfdisplaydoctitle=true,pdfduplex=DuplexFlipLongEdge]{hyperref}
\usepackage{graphicx}  
\usepackage{dcolumn}   
\usepackage{bm}        
\usepackage{amsmath, amsthm, amssymb}
\usepackage{bbold} 
\usepackage{mathrsfs}
\usepackage{array}
\usepackage[usenames]{color}
\usepackage{soul} 
\usepackage{ulem} 
\newcommand{\s}{\sigma}

\newcommand{\phdag}{{\phantom{\dagger}}}

\newcommand{\kv}{\mathbf{k}}

\newcommand{\Qv}{\mathbf{Q}}

\begin{document} 

\normalem

\title{The impact of Rashba spin-orbit coupling in charge-ordered systems} 
\author{Rodrigo A. Fontenele}
\affiliation{Instituto de F\'isica, Universidade Federal do Rio de
Janeiro Cx.P. 68.528, 21941-972 Rio de Janeiro RJ, Brazil}
\author{Sebasti\~ao dos Anjos Sousa J\'unior}
\affiliation{Instituto de F\'isica, Universidade Federal do Rio de
Janeiro Cx.P. 68.528, 21941-972 Rio de Janeiro RJ, Brazil}
\author{Tarik P. Cysne}
\affiliation{Instituto de F\'isica, Universidade Federal Fluminense, 24210-346 Niter\'oi RJ, Brazil}
\author{Natanael C. Costa}
\affiliation{Instituto de F\'isica, Universidade Federal do Rio de
Janeiro Cx.P. 68.528, 21941-972 Rio de Janeiro RJ, Brazil}

\begin{abstract}
We study the impact of the Rashba spin-orbit coupling (RSOC) on the stability of charge-density wave (CDW) in systems with large electron-phonon coupling (EPC).
Here, the EPC is considered in the framework of the Holstein model at the half-filled square lattice.
We start obtaining the phase diagram of the Rashba-Holstein model using the Hartree-Fock mean-field theory, and identifying the boundaries of the CDW and Rashba metal phases.
As our main result, we notice that the RSOC disfavors the CDW phase, driving the system to a correlated Rashba metal. 
Proceeding, we employ a cluster perturbation theory (CPT) approach to investigate the phase diagram beyond the Hartree-Fock approximation.
The quantum correlations captured by CPT indicate that the RSOC is even more detrimental to CDW than previously anticipated.
That is, the Rashba metal region is observed to be expanded in comparison to the mean-field case.
Additionally, we investigate pairing correlations, and the results further strengthen the identification of critical points.
\end{abstract}


\maketitle

\section{Introduction}
 
A growing class of two-dimensional (2D) materials offers opportunities to explore condensed-matter phenomena in a new and flexible platform\,\cite{Balleste2011,GUPTA2015,Novoselov2016,Ahn2020}. In addition to their technological applications, the prospect of solving some old puzzles in solid-state physics motivates much of the fundamental studies in this field. Graphene was the first genuinely 2D material synthesized, but apart from specific situations \cite{Kotov-Uchoa-Netinho-RevModPhys.84.1067}, electronic correlations do not play a significant role on the observed properties, due to their highly dispersive band structure\,\cite{Nath-PhysRevLett.122.077602, Thethe-PhysRevB.72.085123}.
Therefore, part of the research on $2$D materials was developed within the single-particle perspective and neglecting many-body effects\,\cite{CastroNeto-Guinea-RevModPhys.81.109}. With the progress in the quality of samples and synthesis of novel families of materials, a renewed interest in the phases of matter driven by electronic correlation reappeared in the $2$D materials community. Among those compounds of interest, we mention the superconducting phase in twisted bilayer graphene\,\cite{PabloJarilloHerrero-TBG}, the magnetism in Cr$_2$Ge$_2$Te$_6$\,\cite{Nature-magnetism2D-1} and CrI$_3$\,\cite{Nature-magnetism2D-2}, and the charge-density wave (CDW) phase in some transition metal dichalcogenide (TMDs) family\,\cite{Manzeli17}.

For a CDW phase, the key feature is the occurrence of an electron-phonon coupling (EPC), which creates instabilities in the metallic system, leading to a ground state with a spatially-modulated charge order\,\cite{Gruner88,gorkovbook}.
Such a state is characterized by a gapped single-particle charge spectrum and a gapless collective mode formed by phonon excitations.
Interestingly, CDW and superconducting phases have been observed in TMDs compounds which stabilize in the centrosymmetric (1T) \cite{Sugawara-NanoLett-CDW1TTMD} and the non-centrosymmetric (2H) \cite{Lin2020-Natcomm-CDW2HTMD} structures.
In addition, due to the high atomic number of transition metal atoms, the compounds in the TMD family typically have a strong \emph{intrinsic} spin-orbit coupling (SOC).
However, the interplay between the leading effects of a SOC and an EPC, regarding in particular how the SOC affects CDW formation, is still an open issue.
Indeed, the study of the interplay between EPC and SOC in TMDs is challenging due to its complex orbital nature and tri-layer structure, with some works suggesting a SOC-induced renormalization of the electron-phonon coupling\,\cite{Ge2012,Lian2022}.

Different types of spin-orbit interaction may also appear in materials depending on the point-group symmetries of their lattice structure.
In addition to the intrinsic SOC characteristic of materials with high atomic numbers, other types of SOCs can be induced in materials by \emph{extrinsic} mechanisms.
For instance, the Rashba SOC (RSOC) occurs in systems with broken $(z \rightarrow -z)$-symmetry, frequently appearing in surface-confined electrons of semiconductor heterostructures \cite{Takamoto-PhysRevLett.78.1335}, alloys\,\cite{Grioni-PhysRevLett.98.186807, Rotenberg-PhysRevLett.82.4066}, and surface-states of metallic systems\,\cite{Jensen-PhysRevLett.77.3419}.
It can also be induced in 2D materials by proximity to suitable substrates\,\cite{Hongki-PhysRevB.74.165310, Cysne-PhysRevB.98.045407, Cysne-PhysRevB.97.085413}.
In this context, calculations using density functional theory indicate that pristine lead chalcogenides monolayers [Pb$X$, with $X$= S, Se, Te] are 2D materials with strong RSOC\,\cite{CastroNeto-Led-Chalcogenide-PhysRevB.96.161401}.
Its monolayers possess a buckled square lattice that breaks $(z\rightarrow -z)$-symmetry, allowing the occurrence of the Rashba effect, whose intensity is adjusted by changing the buckling angle, via application of strain\,\cite{CastroNeto-Led-Chalcogenide-PhysRevB.97.235312}.
The RSOC may also occur in other 2D materials such as 2D perovskites, and Janus monolayer of transition metals dichalcogenides \cite{2D-SOC-Review-Chen2021}.


One significant impact of the RSOC is the emergence of a spin-dependent momentum shift, resulting in spin-splitting effects in the band structure.
This spin splitting can have profound consequences on the electronic properties of the system, particularly on electron correlations, directly affecting the emergence of new phases.
For instance, in certain families of iridates and pyrochlores -- materials that demonstrate strong electron-electron and spin-orbit interactions -- the interplay between RSOC and electron-electron interactions may give rise to topologically nontrivial states, Weyl semimetals, or even Kitaev spin liquids\,\cite{Pesin10,Krempa14,Rau16}.
Despite recent advancements in this field, the effects of RSOC on \textit{electron-phonon} interacting systems, such as those compounds mentioned earlier, remain less clear.

In view of these stimulating results, in this paper we investigate the impact of RSOC in a charge-ordered system.
Here, we consider dispersionless phonon degrees of freedom coupled to the electrons, within the framework of the Holstein model \cite{Holstein1959}: a simplified model whose ground state at the half-filled square lattice exhibits a staggered CDW phase for any electron-phonon coupling\,\cite{Costa2020}.
This charge-ordered phase may undergo significant changes when external parameters are introduced -- such as electronic doping\,\cite{Dee2019, Bradley2021}, pressure or strain\,\cite{Cohen-Stead2019, Araujo2022}, nonlinear electron-phonon couplings\,\cite{Li2015, Dee2020, Paleari2021}, disorder\,\cite{Xiao2021}, doping with magnetic impurities\,\cite{SASJR2023}, or phonon dispersion\,\cite{Costa2018} --, with the enhancement of pairing correlations for most of cases.
Despite this scrutinization, the impact of a SOC on the Holstein CDW phase is barely unknown.
We bridge this gap by examining the Rashba-Holstein Hamiltonian through two methodologies: mean-field (MF) and cluster perturbation theory (CPT) approaches.

The paper is organized as follows: we present the Rashba-Holstein model in Sec.\,\ref{sec:Model}, with the results for the MF and CPT approaches being discussed in Secs.\,\ref{subsec:MFT} and \ref{subsec:CPT}, respectively.
In Section \ref{sec:conc} we present our conclusions.  


\section{The Rashba-Holstein Hamiltonian \label{sec:Model}}

We consider electrons hopping in a single orbital square lattice.
Here, we assume the system has a broken $(z\rightarrow -z)-$symmetry, as occurs in lead chalcogenides monolayers \cite{CastroNeto-Led-Chalcogenide-PhysRevB.96.161401, CastroNeto-Led-Chalcogenide-PhysRevB.97.235312} and in confined electron gases \cite{Takamoto-PhysRevLett.78.1335, Grioni-PhysRevLett.98.186807, Rotenberg-PhysRevLett.82.4066, Jensen-PhysRevLett.77.3419}, which allows the occurrence of RSOC.
In addition, we use the Holstein model to describe the electron-phonon coupling, which captures the local interaction of electrons with single-Einstein phonon degrees of freedom\,\cite{Holstein1959}.

For the Rashba-Holstein model, the Hamiltonian reads
\begin{eqnarray}
\mathcal{H} =&& -t \sum_{\langle \textbf{i},\textbf{j} \rangle,\sigma}( c_{\textbf{i},\sigma}^\dagger c_{\textbf{j},\sigma} + h.c. )- \mu \sum_{\textbf{i},\sigma} n_{\textbf{i},\sigma} \nonumber \\
    && +\hbar \omega_{0}\sum_{\textbf{i}} a_{\textbf{i}}^\dagger a_{\textbf{i}}
    -   g  \sum_{\textbf{i},\sigma}  n_{\textbf{i},\sigma} ( a_{\textbf{i}} + a_{\textbf{i}}^\dagger  ) \nonumber \\
    && -t_{R}\sum_{\langle \textbf{i},\textbf{j} \rangle}c_{\textbf{i}}^{\dagger}\left( {\overleftrightarrow \alpha}_{\textbf{i},\textbf{j}} \times {\overleftrightarrow \sigma}\right)_{z}c_{\textbf{j}}~,
\label{eq:HHR}
\end{eqnarray}
where the sums run over a square lattice, with $\langle \textbf{i},\textbf{j} \rangle$ denoting nearest neighbors.
Here we use the usual second quantization formalism, with $c_{\textbf{i},\sigma}$ ($c_{\textbf{i},\sigma}^{\dagger}$) being a fermionic operator that annihilates (creates) an electron with spin $\sigma=\uparrow, \downarrow$ on a given site $\mathbf{i}$; $n_{\textbf{i},\sigma}=c_{\textbf{i},\sigma}^{\dagger} c_{\textbf{i},\sigma}$ is the fermion number operator.
Similarly, $a_{\textbf{i}}$ ($a_{\textbf{i}}^{\dagger}$) is a bosonic operator that annihilates (creates) a phonon with energy $\hbar \omega_{0}$ on a given site $\mathbf{i}$.
In the last term of Eq. (\ref{eq:HHR}), ${\overleftrightarrow \alpha}_{\textbf{i},\textbf{j}} \equiv \left(\alpha_{\textbf{i},\textbf{j}}^{x}, \alpha_{\textbf{i},\textbf{j}}^{y}, 0 \right)$ is the displacement tensor $\alpha_{\textbf{i},\textbf{j}}^{\nu} \equiv i(\delta_{\textbf{i},\textbf{j}+\alpha_{\nu}}-\delta_{\textbf{i},\textbf{j}-\alpha_{\nu}})$, with $\alpha_{\nu}$ being unitary translation in the $\nu$ direction, and ${\overleftrightarrow \sigma} \equiv \left( \sigma_{x}, \sigma_{y}, \sigma_{z}\right)$ are Pauli matrices related to electronic spin.
The first two terms on the right-hand side of Eq.\,\eqref{eq:HHR} correspond to the hopping of the electrons and their chemical potential, respectively.
The third term describes the local harmonic oscillators, while the fourth represents the EPC, with strength $g$.
Finally, the last term is the RSOC, with strength $t_R$.
We set the energy scale in units of the hopping integral $t$.

At this point, it is also worth mentioning that the ground state of the half-filled pure Holstein model ($t_{R}=0$) is CDW for any EPC $g > 0$, due to instabilities from the van Hove singularity (vHs) and the Fermi surface nesting (FSN)\,\cite{Hohenadler19,Costa2020}.
On the other hand, for a pure Rashba metal, i.e.~$t_{R} \neq 0$ and $g = 0$, both FSN and vHs are absent at the half-filling, due to the splitting of the spin textures -- similar to those presented in Fig.\,\ref{fig:fig1}\,(a) and (c).
Such feature leads to no divergence in the density of states at the Fermi level $N(\omega=0)$ -- consequently, no divergence in the noninteracting electronic susceptibility.
Therefore, within a weak coupling analysis of these two cases, one should expect that the RSOC must suppress CDW formation.
Although this may be true for the weak coupling case, the scenario is less clear in the intermediate and strong coupling regimes, where the Fermi surface does not play a relevant role.
In what follows, we investigate this problem at the ground state, by performing analyses within both MF and CPT approaches, fixing $\omega_{0}/t = 1$ while varying $t_R$ and $g$.
  

\section{Mean-field theory \label{subsec:MFT}} 

We start by discussing the mean-field approach to the Rashba-Holstein Hamiltonian.
To this end, we change the notation of Eq.\,\eqref{eq:HHR}, rewriting the phonon creation and annihilation operators in terms of position and momentum ones.
For instance, by considering the pure Holstein model ($t_R = 0$), its Hamiltonian is
\begin{eqnarray}
    \mathcal{H}_{\text{H}} =&& -t \sum_{\langle \textbf{i,j} \rangle,\sigma}( c_{\textbf{i},\sigma}^\dagger c_{\textbf{j},\sigma} + h.c. )- \mu \sum_{\textbf{i},\sigma} n_{\textbf{i},\sigma} \nonumber \\
    && -g \sqrt{\frac{2M\omega_{0}}{\hbar}}  \sum_{\textbf{i},\sigma}  n_{\textbf{i}}^\sigma \hat{X}_{\textbf{i}}+ \sum_{\textbf{i}}  \left( \dfrac{\hat{P}_{\textbf{i}}^2}{2M} + \dfrac{M \omega^2_0}{2} \hat{X}_{\textbf{i}}^2   \right), \nonumber \\
\end{eqnarray}
with $\hat{X}_{\bf{i}} = \sqrt{ \dfrac{\hbar}{2M\omega_0} } (a_\textbf{i} + a_{\textbf{i}}^\dagger)$ and $\hat{P}_{\bf{i}} = \mathit{i}\sqrt{ \dfrac{\hbar M \omega_0}{2} } (a_{\textbf{i}}^\dagger - a_\textbf{i} )$ being the position and momentum operators, respectively, of a harmonic oscillator in a given site $\mathbf{i}$.
Here, we define the mass of the harmonic oscillators ($M$) and the lattice constant as unities.

Within a static mean-field approach, we assume that the kinetic energy of lattice vibrations is negligible if compared with potential energy ($\hat{P}_{\textbf{i}}^2/2M \ll M \omega_0^2 \hat{X}_{\textbf{i}}^2/2$), i.e., we deal with a permanent distortion that breaks the translation symmetry of the lattice.
This assumption corresponds to the adiabatic regime \cite{adiabatic-limit-PhysRevB.81.115114, adiabatic-limit-PhysRevB.49.9915}, i.e.~$\omega_0 \to 0$ while keeping $M \omega^{2}_0 \equiv K$ finite.
In this regime, the Hamiltonian is a quadratic form for electron creation/annihilation operators, therefore its ground state is exact by a mean-field approach.

Proceeding, we approximate the position operator by its expected value, assuming the \textit{ansatz} 
$\hat{X}_\textbf{i} \rightarrow  \langle X_{\bf{i}} \rangle = X_0 + \cos ({\bf{Q}} \cdot {\bf{i}}) X_1$,
with $\textbf{Q} = (\pi,\pi)$.
Notice that other ansatze involving arbitrary wavevectors $\textbf{Q} = (q_x, q_y)$ can be explored, even in incommensurate cases\,\cite{Lopes23}.
However, owing to the four-fold rotational symmetry inherent to the square lattice, along with the nesting properties and the commensurate filling, it is expected that the $(\pi, \pi)$ configuration is the one that most significantly reduces the internal energy.
Therefore, the MF Holstein Hamiltonian becomes
\begin{eqnarray}\label{eq:hamil_momentum}
    \mathcal{H}_{\text{H}} = && -t \sum_{\langle \textbf{i,j} \rangle,\sigma}( c_{\textbf{i},\sigma}^\dagger c_{\textbf{j},\sigma} + \text{H.c.} )- \mu \sum_{\textbf{i},\sigma} n_{\textbf{i},\sigma} \nonumber \\ 
    && -g\sqrt{\frac{2M\omega_0}{\hbar}}  \sum_{\textbf{i},\sigma}  n_{\textbf{i},\sigma} \big[ X_0 +  X_1 \cos ({\bf{Q}} \cdot {\bf{i}}) \big]  \nonumber \\ 
    && + \dfrac{N M \omega^2_0}{2} \left( X_0^2 + X_1^2   \right),
\end{eqnarray}
where $N$ is the number of lattice sites.

\begin{figure}[t]
	\centering
    \includegraphics[scale=0.5]{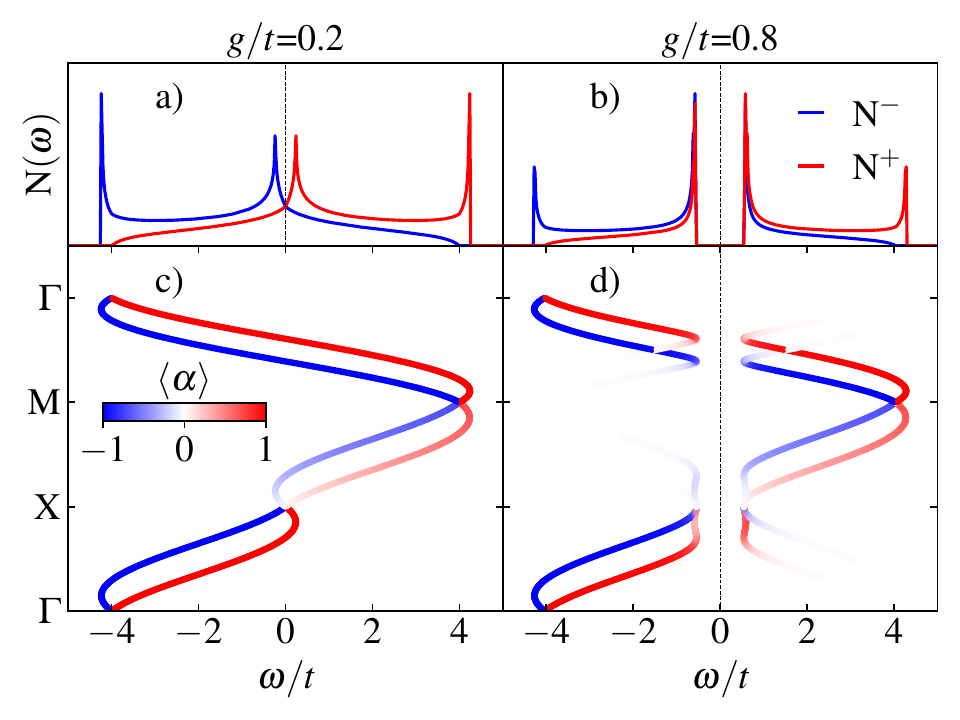}   
	\caption{The mean-field result for the density of states of the Rashba-Holstein model for EPC (a) $g/t = 0.2$, and (b) $g/t = 0.8$, and its electronic dispersion in (c) and (d), respectively. The results are for fixed RSOC $t_{\rm R}/t=0.5$, with the color codes indicating the chiral spin-texture polarization of bands.}
	\label{fig:fig1}
\end{figure}

The inclusion of the Rashba term changes the electron dispersion, and the eventual mean-field Hamiltonian in reciprocal space reads
\begin{eqnarray}
	\mathcal{H}_\text{MF} &&= \sum_{\kv,\s}
	(\varepsilon_\kv - \mu^\prime ) c_{\kv\s}^\dagger c_{\kv\s}^\phdag + \sum _{\kv,\s,\s'} (\widehat{V}_\kv)_{\s\s'}^\phdag c_{\kv\s}^\dagger c_{\kv\s'}^\phdag,   \nonumber \\ 
    &&-g\sqrt{\frac{M\omega_0}{2\hbar}} X_1 \sum_{\kv,\s} (c_{\kv+\Qv\s}^\dagger c_{\kv\s}^\phdag + \text{H.c.}) \nonumber \\
    &&+ \dfrac{N M \omega^2_0}{2} \left( X_0^2 + X_1^2   \right),
\label{eq:hamkspace}
\end{eqnarray}
with $\varepsilon _\kv = -2t(\cos{k_x}+\cos{k_y})$, $\widehat{V}_\kv \equiv 2t_\text{R}(\s_x\sin{k_y} - \s_y\sin{k_x})$, and $\mu^\prime = \mu + g\sqrt{2M\omega_0/\hbar} X_0 $.
In a basis of Nambu spinors $\Phi^{\dagger}_\kv = ~(c_{\kv\uparrow}^\dagger, c_{\kv\downarrow}^\dagger,  c_{\kv+\Qv\uparrow}^\dagger, c_{\kv+\Qv\downarrow}^\dagger)$, we obtain a quadratic form representation of $\mathcal{H}_{MF}$,
\begin{eqnarray}
\mathcal{H}_\text{MF}  = \frac{1}{2}\sum _\kv \Phi^{\dagger}_\kv \widehat{H}_\kv \Phi _\kv   +  \dfrac{N M \omega^2_0}{2} \left( X_0^2 + X_1^2   \right)
\label{eq:spinbasis}
\end{eqnarray}
where
\begin{eqnarray}
	 \widehat{H}_\kv = \left(
		\begin{array}{cccc}
\varepsilon_{\kv} - \mu^\prime & (\widehat{V}_\kv)_{ \uparrow \downarrow} & - \Delta X & 0\\
(\widehat{V}_\kv)_{\downarrow \uparrow} & \varepsilon_{\kv} - \mu^\prime & 0  & - \Delta X \\
- \Delta X & 0 & \varepsilon_{\kv+\Qv} - \mu^\prime  & (\widehat{V}_{\kv+\Qv})_{\uparrow \downarrow} \\
0 & - \Delta X & (\widehat{V}_{\kv+\Qv})_{\uparrow \downarrow} & \varepsilon_{\kv+\Qv} - \mu^\prime \\
\end{array}
\right)
\label{eq:HMF}
\end{eqnarray}
with $\Delta X =  g\sqrt{2 M\omega_0/\hbar} X_1 $.
Notice that we added $X_{0}$ to the chemical potential $\mu^\prime$, since this field shifts the energy level, while the $X_1$ field modulates the electronic distribution, therefore being proportional to the amplitude of the CDW order parameter.
\footnote{Indeed, if we define $\langle n_{i} \rangle = n + \cos ({\bf{Q}} \cdot {\bf{i}}) \delta n$, from Eq.\,\eqref{eq:hamil_momentum} one obtains $\delta n = \omega_0 \sqrt{\frac{M\hbar\omega_0}{2}} \frac{X_1}{g}$.}

\begin{figure}[t]
\includegraphics[scale=0.5]{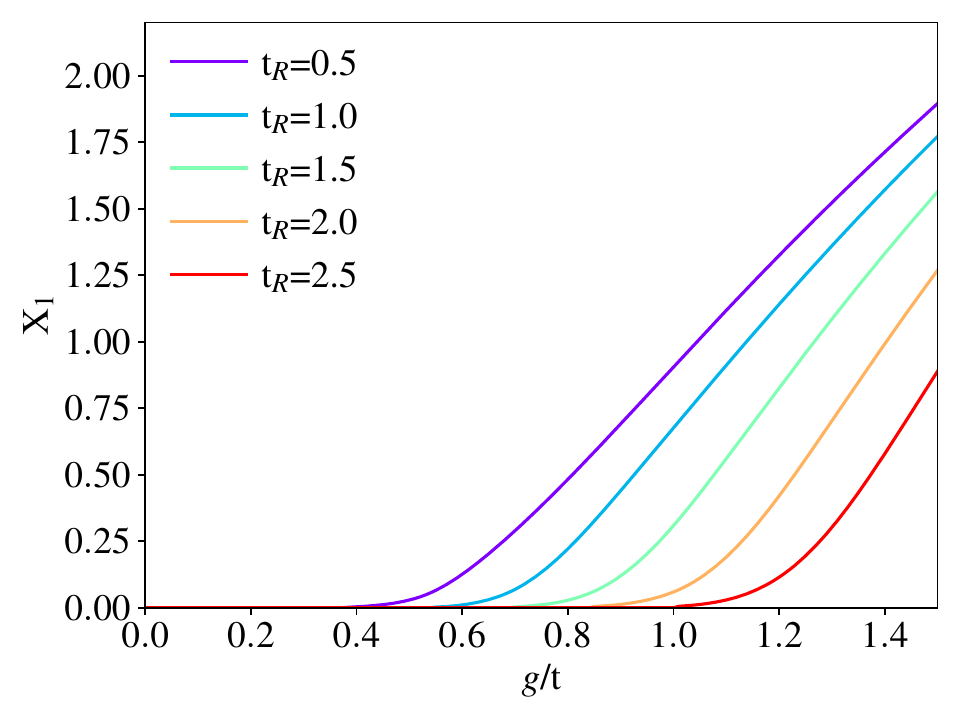} 
\caption{The behavior of the ground state mean-field parameter $X_{1}$ (proportional to the CDW order parameter) as a function of the electron-phonon coupling $g$, for different values of RSOC $t_R$'s.}
\label{fig:order_parameter} 
\end{figure}

The parameters $X_{0 (1)}$ are determined self-consistently by minimizing the Helmholtz free energy, through the diagonalization of Eq.\,\eqref{eq:spinbasis}.
We use the minimization solutions to describe the ground state properties of the system, such as the density of state (DOS) and the electron dispersion, e.g., as displayed in Fig.\,\ref{fig:fig1} for small and large values of EPC, while keeping $t_R/t=0.5$ fixed.
As presented in Figs.\,\ref{fig:fig1}\,(a) and (c), when $g/t=0.2$, the RSOC splits the electronic spectra into two branches with well-defined chiralities, $\langle \overleftrightarrow \sigma \rangle. ({\bf k}\times \hat{z})/|{\bf k}|=\pm 1$ for right-handed and left-handed.
Such a split shifts the energy levels and, consequently, the van Hove singularities, leading to a reduction in the DOS at the half-filling, as shown in Fig.\,\ref{fig:fig1}\,(a).
On the other hand, for large EPC, a CDW phase emerges and creates a charge gap at the half-filling, as displayed in Figs.\,\ref{fig:fig1}\,(b) and (d), for ~$g/t=0.8$.
This result provides a few hints that the RSOC is harmful to the CDW phase.

To further emphasize the effects of the RSOC on the CDW phase, Fig.\,\ref{fig:order_parameter} displays the behavior of $X_1$ as a function of $g/t$, for different spin-orbit couplings, in the ground state.
For fixed values of $g/t$, where the minimization of the energy leads to CDW, it is clear that the main effect of the RSOC is to reduce $X_1$, which is deleterious to charge modulation.
At this point, a remark should be made: the asymptotic behavior to $X_1$ as a function of $g/t$ may be an artifact of the mean-field solution.
Then, we define $X_1 = 10^{-3}$ as the lower bound limit for the CDW phase, which is enough to reproduce $g_c \to 0$ when $t_R \to 0$, as expected for the pure Holstein model at the half-filled square lattice.
This lower bound limit defines the critical region where the CDW phase is destroyed (or emerges).

\begin{figure}[t]
    \centering
    \includegraphics[scale=0.5]{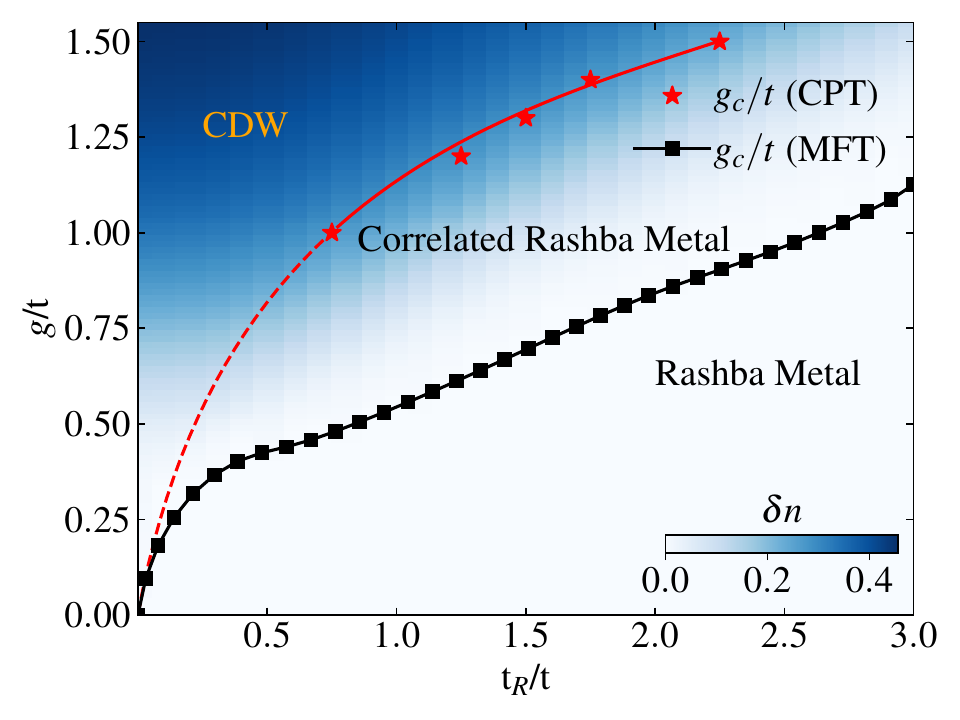}
    \caption{The ground-state phase diagram of Holstein-Rashba model in the half-filled square lattice. The black square symbols correspond to the mean-field solution, while the red star ones denote the CPT results. The blue heatmap indicates the amplitude of charge modulation in the CDW phase for the MFT solution. The red curve is just a guide to the eye.}
    \label{phasediagram}
\end{figure}

\begin{figure}[t]
    \centering
    \includegraphics[scale=0.5]{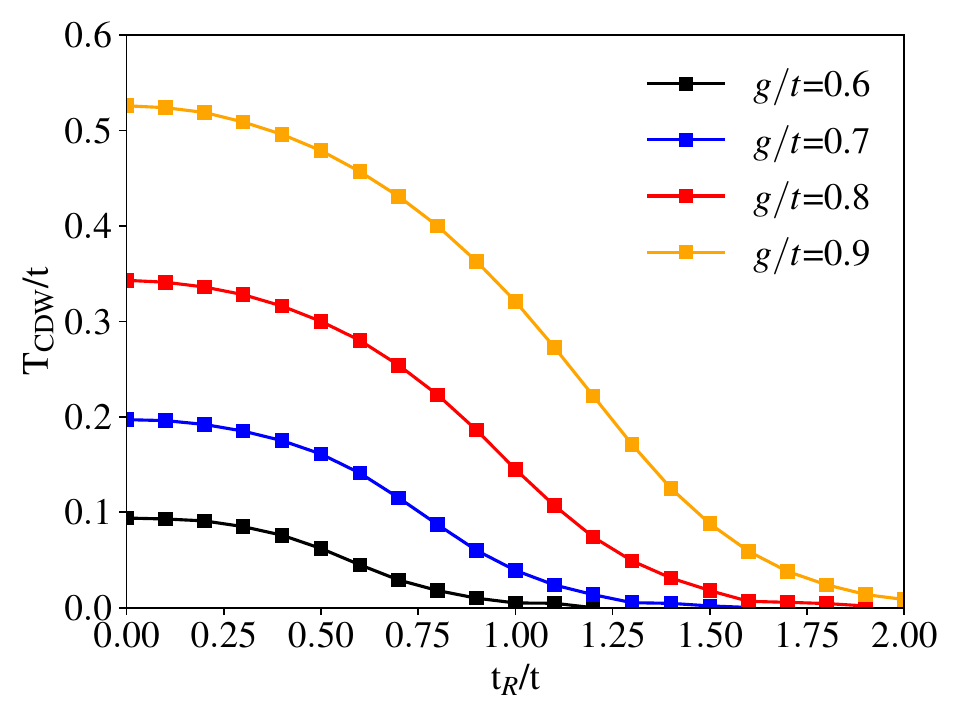}
    \caption{CDW critical temperature as a function of $t_R/t$ for different values of $g/t$.}
    \label{fig:tcs}
\end{figure}

Repeating similar analyses as those of Fig.\,\ref{fig:order_parameter}, we are able to present the mean-field phase diagram of the Rashba-Holstein model at half-filling, as shown in Fig.\,\ref{phasediagram}.
Here, the CDW and Rashba metal phases are separated by the curve going through the black squares, while the color-map scheme describes the amplitude of the CDW charge modulation, $\delta n$.
As complementary information, Fig.\,\ref{fig:tcs} displays the critical CDW temperature $T_{\rm CDW}$, below which the system opens a gap\,\footnote{Again, notice that tail-like behavior of $T_{\rm CDW}$ for large SOC is a mean-field artifact.}.
In line with the previous discussion, the RSOC tends to suppress the CDW phase, decreasing both the critical temperatures and the order parameter, eventually leading to a Rashba metal.

Despite being obtained by a static mean-field approach, the phase diagram of Fig.\,\ref{phasediagram} is expected to provide the correct transition lines when $\omega_0 \to 0$.
However, $\omega_{0}/t = 1$ is clearly away from this limit, which may require less biased methodologies to investigate the Hamiltonian properties, as discussed in the next Section.

\section{Cluster Perturbation Theory \label{subsec:CPT}}

Now, we use the cluster perturbation theory, which is a convenient technique to extract spectral weights and the electronic DOS of a correlated system. Briefly speaking, the CPT consists of partitioning an infinite lattice into clusters of finite size where the Hamiltonian is exactly solved, while coupling the clusters by perturbation theory approaches.
A pedagogical introduction to this method can be found in Ref.\,\onlinecite{CPT-Holstein-PhysRevB.68.184304,CPT-Hubbard-PhysRevB.66.075129, CPT-Hubbard-PhysRevLett.84.522, senechal2010introduction}.

Within this approach, we decompose the Hamiltonian as
\begin{align}
    H = H_{\text{cluster}} + T~,
    \label{eq:Hcpt}
\end{align}
where
$H_{\text{cluster}}= K + H_{\text{int}}$
describes the operations inside a given finite-sized cluster, with $K = \sum_{\alpha, \beta} t_{\alpha,\beta}c_{\alpha}^{\dagger}c_{\beta}$ and $H_{\text{int}} = -g  \sum_{\textbf{i},\sigma}  n_{\textbf{i},\sigma} ( a_{\textbf{i}} + a_{\textbf{i}}^\dagger  )$ being the hopping and the EPC terms, respectively. 
The term $T = \sum_{\alpha, \beta}V_{\alpha, \beta}c_{\alpha}^{\dagger}c_{\beta} $ is the one that connects different clusters of the superlattice (i.e., intercluster sites).
Given this, one can write the Green's functions as
\begin{align}
    \textbf{G}(\omega)^{-1} = \mathbf{G}'(\omega)^{-1} - \mathbf{V},
    \label{eq:TFGFcpt1}
\end{align}
where $\textbf{G}(\omega)$ and $\mathbf{G}'(\omega)$ are the infinite lattice and the (exact) finite-sized cluster Green's functions, respectively.
As the cluster superlattice has a reduced Brillouin zone, one can connect any wavevector ${\bf k}$ of the original Brillouin zone to a wavevector ${\bf \tilde{k}}$ in the reduced one by ${\bf k}={\bf \tilde{k}}+{\bf K}$, where ${\bf K}$ is an element of the reciprocal superlattice. Therefore, in the reduced Brillouin zone, Eq.\,\eqref{eq:TFGFcpt1} is given by
\begin{align}
    \textbf{G}(\textbf{\~k},\omega)^{-1} = \mathbf{G}'(\omega)^{-1} - \mathbf{V}(\textbf{\~k})~,
    \label{eq:TFGFcpt2}
\end{align}
which eventually leads to the CPT Green's functions
\begin{align}
    G_{\text{CPT}}(\mathbf{k},\omega) = \frac{1}{L}\sum_{a,b}e^{-i\mathbf{k}(\mathbf{r}_{a} - \mathbf{r}_{b})}G_{a,b}(\mathbf{\tilde{k}},\omega)~,
    \label{eq:GreenCPT}
\end{align}
with ${\bf r}_{a(b)}$ being the position of the intra-cluster sites.

The CPT Green's functions take into account the leading order in the interaction term, and allow one to obtain the spectral weight
\begin{align}
    A_{0}(\mathbf{k},\omega) = -\frac{1}{2\pi}\text{Im}[G_{\text{CPT}}(\mathbf{k},\omega)],
    \label{eq:A0}
\end{align}
and the DOS
\begin{align}
    N(\omega) = \frac{1}{N}\sum_{\mathbf{k}} A_{0}(\mathbf{k},\omega).
    \label{eq:DOS}
\end{align}
In what follows, we use the DOS to identify the transition from a gapped CDW phase to a Rashba metal. 

\begin{figure}[t]
\centering
\includegraphics[scale=0.5]{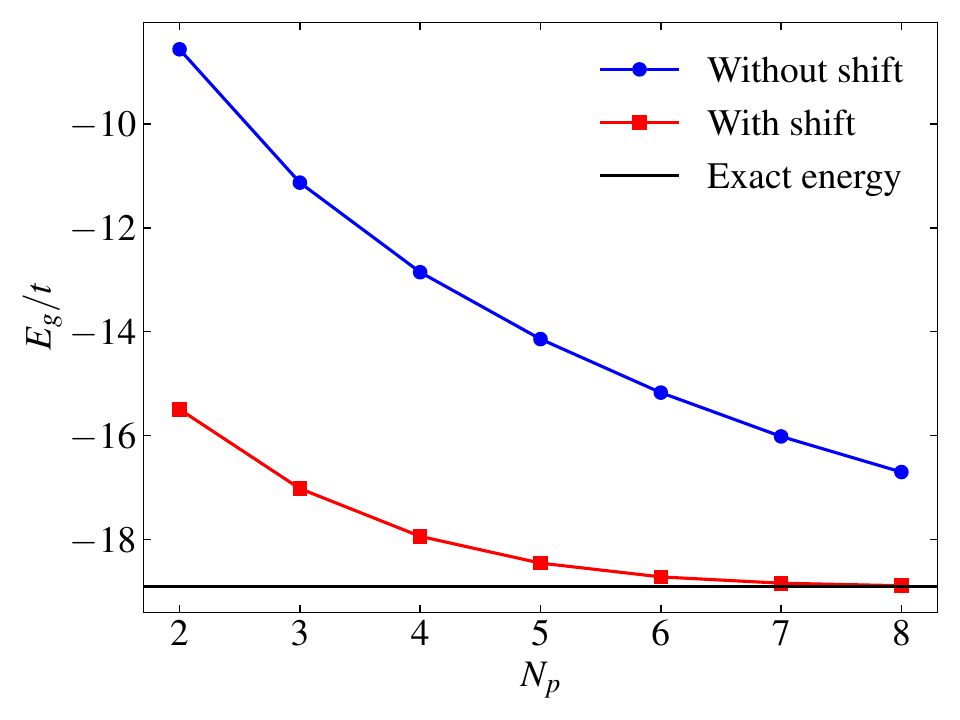} 
\caption{The ground state energy of the Holstein model for $g/t=1.5$, in a $2\times 2$ plaquette, as a function of the cutoff number of phonons per site $N_{p}$. The red squares (blue circles) symbols denote the ground state energies for the case with (without) a shift to symmetrize the electron-phonon term.}
\label{fig:cutoff} 
\end{figure}

At this point, it is worth discussing a few technical issues about our CPT implementation.
Due to the phonon degrees of freedom, the Hilbert space is infinite, which demands a cutoff in the total number of phonons to perform the exact diagonalization procedure.
Among the different ways of establishing this cutoff, we have defined a maximum limit, $N_{p}$, for the number of phonons per site.
Its value is determined through a systematic analysis of the ground state energy $E_g$ as a function of $N_{p}$, as displayed in Fig.\,\ref{fig:cutoff}, for fixed $g/t= 1.5$ and $\omega_{0}/t=1$, in a $2\times 2$ plaquette.
Notice that $E_g(N_{p})$ goes asymptotically to the limit $N_{p} \to \infty$, so the error from the cut-off may be disregarded for a given value of $N_{p}$.
However, the electron-phonon term described in Eq.\,\eqref{eq:HHR} requires a substantial number of phonons to attain the correct ground state, as evidenced by the behavior of the blue circle symbols in Fig.\,\ref{fig:cutoff}.
To circumvent this difficulty and keep $N_{p}$ as small as possible, we perform a shift in the phonon operators $a_{\mathbf{i}} \to \big( a_{\mathbf{i}} + g \langle n \rangle \big)$, with $ \langle n \rangle$ being the average electron density.
We emphasize that such a shift is equivalent to a change in the origin of the phonon position operators $\hat{X}_{\mathbf{i}}$ to deal just with excitations around its mean value.
This drastically reduces the number of required phonons, as shown by the behavior of the red square symbols in Fig.\,\ref{fig:cutoff}.
Despite this improvement, the greater the EPC, the larger the values of $N_{p}$, therefore we limit our analyses to intermediate coupling strengths for clusters of $2\times 2$ sites.

\begin{figure}[t]
\includegraphics[scale=0.5]{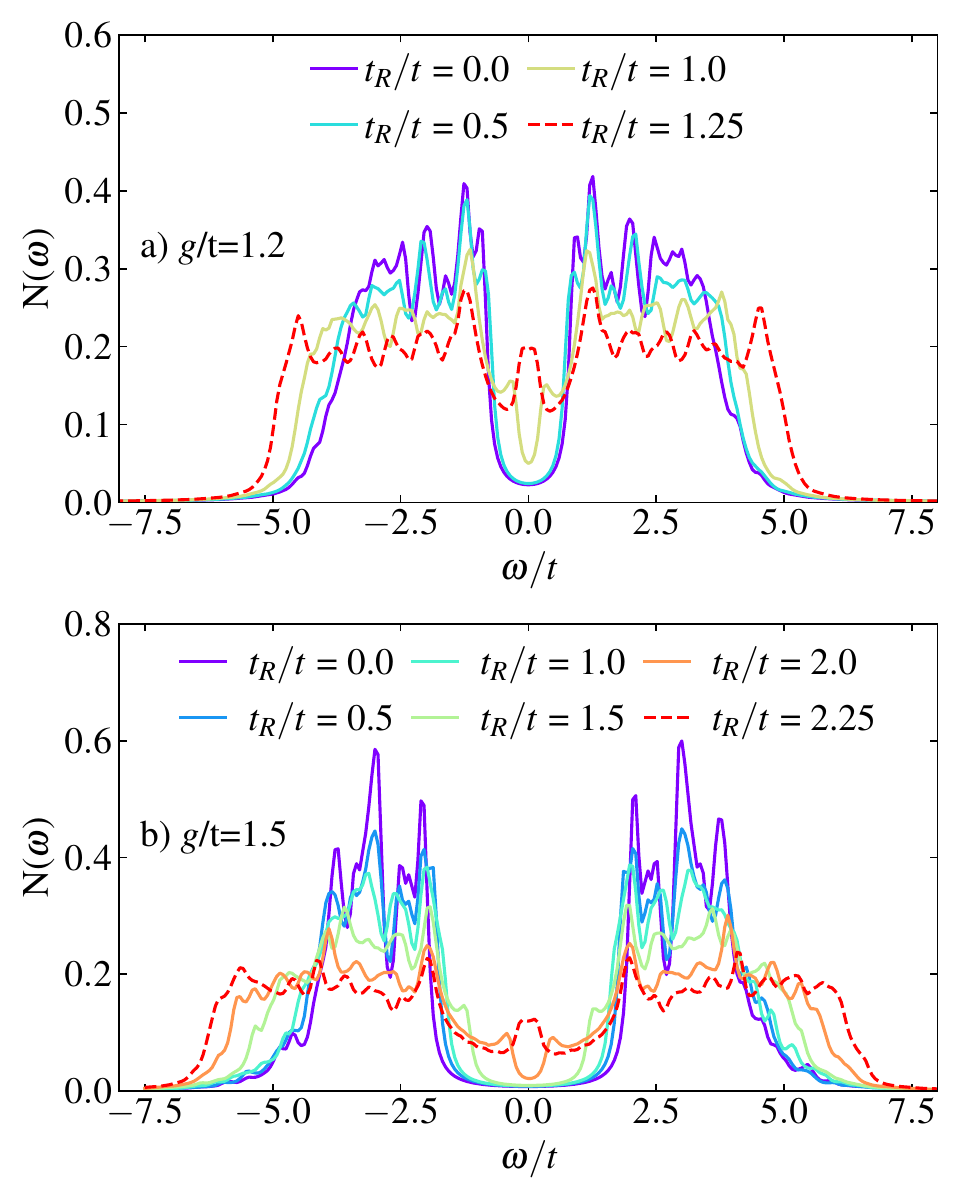}
\caption{The DOS of the Rashba-Holstein model for EPC (a) $g/t=1.2$ and (b) $g/t=1.5$, for different values of RSOC $t_{R}$.}
\label{fig:DOS_plaquete_lamb1.2} 
\end{figure}

Now, using the CPT method we turn to discuss the effects of the RSOC on the spectral properties of our charge-ordered system, for fixed $\omega_{0}/t=1$, and at half-filling.
Figure \ref{fig:DOS_plaquete_lamb1.2}\,(a) displays the behavior of $N(\omega)$ for $g / t = 1.2$ and different values of $t_R$.
When $t_R=0$, the system has a large charge gap around $\omega=0$, in line with our expectation of a CDW phase for the Holstein model in the half-filled square lattice.
Interestingly, the size of the gap remains nearly unchanged for $t_R/t \lesssim 0.5$, which shows that the CDW phase is robust against spin-orbit perturbations.
However, for larger RSOC, the charge gap drastically decreases, until $N(\omega \approx 0)$ exhibits a peak instead of a gap for $t_R/t \approx 1.25$.
This specific change in the behavior of $N(\omega)$ around $\omega \approx 0$ (and $t_R/t \approx 1.25$) is a clear indication of the occurrence of a phase transition from the CDW phase to a Rashba metal.

A very similar behavior occurs for different values of $g$, as displayed in Fig.\,\ref{fig:DOS_plaquete_lamb1.2}\,(b), for $g / t = 1.5$.
The main difference between this case and the previous one is the need for a larger RSOC to destroy the charge gap.
In view of this, here we define as \textit{critical points} the values of $t_R$ at which $N(\omega)$ exhibits a peak instead of a gap at the half-filling.
These CPT critical points are presented as red star symbols in the phase diagram of Fig.\,\ref{phasediagram}, with the red (dashed and solid) lines being just guides to the eye.

\begin{figure}[t]
\includegraphics[scale=0.5]{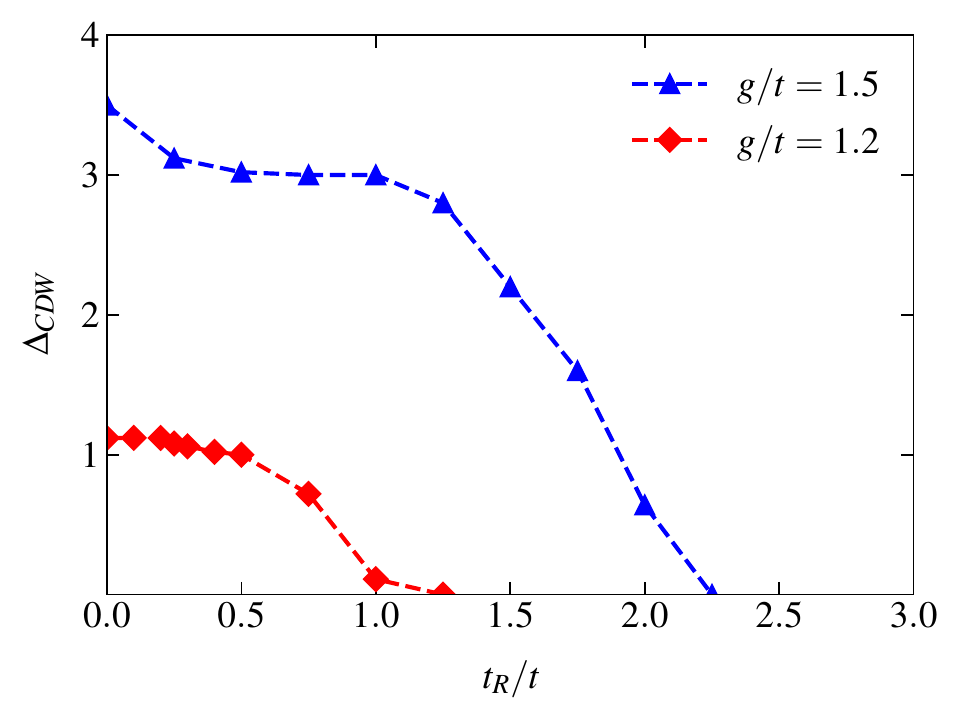}
\caption{The CDW charge gap, $\Delta_{\rm CDW}$ as a function of the RSOC, $t_R/t$, for fixed $\omega_{0}/t=1$.}
\label{fig:CDW_Gap} 
\end{figure}

In order to further emphasize the phase transitions, we extract the charge gap directly from Fig.\,\ref{fig:DOS_plaquete_lamb1.2} by defining the CDW order parameter, $\Delta_{\rm CDW}$, as the energy difference between the two peaks away from $\omega =0$.
The behavior of $\Delta_{\rm CDW}$ as a function of $t_R$ is displayed in Fig.\,\ref{fig:CDW_Gap}, for $g / t =1.2$ (red diamond symbols) and 1.5 (blue triangle symbols), from which one may notice that the gap closing is consistent with a second order phase transition.
In particular, the suppression of the CDW phase seems to occur at $t_{R(c)} = 1.1 \pm 0.1$ for $g / t =1.2$, and at $t_{R(c)} = 2.1 \pm 0.1$ for $g / t =1.5$.

There are clear differences between the CPT and MFT results, which may be understood through two key remarks.
First, and foremost, the CPT method takes into account the effects of short-range electronic correlations, while the MFT deals with local ones.
That is, the MFT exhibits the points where the Rashba metal becomes on-site correlated, while the CPT provides insights when these correlations become long-ranged.
Therefore, the region enclosed between the MFT and CPT lines in Fig.\,\ref{phasediagram} defines a \textit{correlated} Rashba metal, i.e.~metallic (from CPT) but with on-site correlations (from MFT approach).
As a second remark, we recall that the mean-field solution is exact at the ground state when $\omega_{0} \to 0$, thus the MFT line in Fig.\,\ref{phasediagram} can also be interpreted as as the boundary of the adiabatic limit.
That is, if one reduces $\omega_{0}$, the CPT transition line must be pushed into the MFT ones.
Then, it provides the behavior of the critical points when $\omega_{0}$ changes.

\begin{figure}[t]
\centering
\includegraphics[scale=0.5]{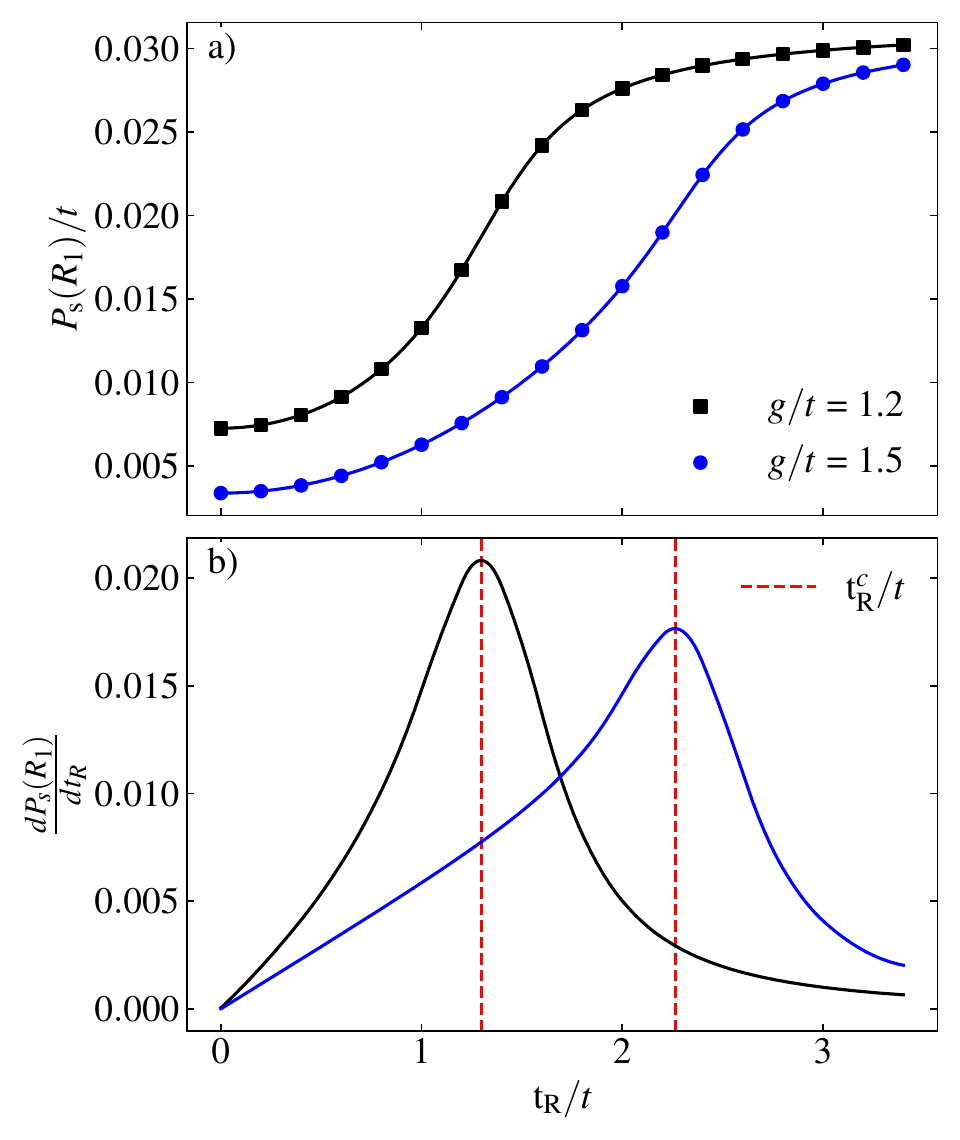}
\caption{The s-wave pair correlation function $P_s({\bf i}, {\bf j})$ as a function of RSOC and its numerical derivative $\text{d} P_s({\bf i}, {\bf j})/ \text{d} t_{\rm R}$ for two values of EPC. The correlation function is calculated between pairs at a site ${\bf i}$ and its nearest-neighbor ${\bf j}$ symbolized here by $({\bf i}, {\bf j})=(R_1)$.}
\label{fig:PsR1_and_derivate} 
\end{figure}

At this point, it is worth discussing the emergence of pairing correlations in the absence of the CDW ones.
As the occurrence of superconductivity is a challenge for both methodologies, we analyze the \textit{s}-wave pair correlation functions\,\cite{Paiva04}
\begin{eqnarray}
P_s({\bf i}, {\bf j})=\frac{1}{2} \langle b^{\dagger}_{{\bf i}} b_{{\bf j}}+\text{H.c.} \rangle
\label{Eq:Cij}
\end{eqnarray}
with $b^{\dagger}_{{\bf i}}=c^{\dagger}_{{\bf i}\uparrow} c^{\dagger}_{{\bf i}\downarrow}$ ($b_{{\bf j}}=c_{{\bf j}\downarrow} c_{{\bf j}\uparrow}$), restricting our analysis to nearest-neighbor sites.
This function is computed in real space, after partitioning the system in the cluster scheme described above.
Figure \ref{fig:PsR1_and_derivate}\,(a) presents $P_s(|{\bf i}-{\bf j}|=1)$ as a function of $t_{\rm R}$ for $g / t = 1.2$ and 1.5, from which we see an enhancement of the pairing correlations.
This result provides us with an interesting feature: the derivative of the nearest-neighbor pair correlation function ($d P_s/d t_R$) exhibits peaks for values very close to the CDW critical points, as shown in Fig.\,\ref{fig:PsR1_and_derivate}\,(b).
Such a behavior is not a coincidence; as charge-charge correlations compete with pairing ones, then it is expected that $P_s({\bf i},{\bf j})$ should be enhanced around the points where the CDW gap closes, independently whether the ground state is superconducting or not.
Therefore, although this result is not enough to support the existence of superconductivity, it reinforces our previous findings about the nature of the charge gap being a CDW phase, and about the location of the CPT critical points presented in the phase diagram of Fig.\,\ref{phasediagram}. 

Last, but not least, we recall that the spin-orbit coupling has been investigated in different contexts of strongly correlated systems, in particular for those involving electron-electron interactions.
Then, it is imperative to compare these cases with ours, highlighting both their similarities and differences.
For instance, for the repulsive Hubbard model in the half-filled square lattice, the inclusion of the RSOC leads to the emergence of novel magnetic phases, which may include spiral and striped ones; away from the half-filling, ferromagnetism may appear\,\cite{Zhang2015, Brosco2020, Kennedy2022, WANG2023, Kawano2023, Hodt2023}.
This occurs due to the spin-flip term in the RSOC, which disfavors the staggered antiferromagnetic ground state of the half-filling, suppressing the charge gap, while keeping gapless spin excitations.
Due to the spin dependence in the interaction term in the Hubbard model, the CPT spectral properties for this case exhibit a clear dependence of the chiral bands and DOS as a function of the Hubbard interaction strength $U$\,\cite{Brosco2020}.
It is substantially different from our present study, since the interaction term in the Rashba-Holstein model is not spin-dependent, with the spin and charge gaps emerging/disappearing together.
Furthermore, noteworthy changes manifest in the attractive Hubbard model when subjected to the RSOC.
This model exhibits \textit{s}-wave superconductivity for any filling\,\cite{Paiva04,Moreo91,Fontenele22}, whose critical temperatures are significantly increased with the inclusion of the RSOC\,\cite{Tang2014, Rosenberg2017, Ptok2018}.
This enhancement results from the suppression of charge-charge correlations, in line with our findings for the Rashba-Holstein model.
Given that the antiadiabatic regime ($\omega_{0} \to \infty$) of the Holstein one can be mapped onto the attractive Hubbard model, we anticipate the emergence of superconductivity in our scenario.
However, achieving this would require the application of unbiased methodologies, such as quantum Monte Carlo methods, which is beyond the scope of the present study.


\section{Conclusions} 
\label{sec:conc}

In this work, we have studied the impact of the Rashba spin-orbit coupling to a charge-ordered system, as the Holstein model.
To this end, we investigated the Rashba-Holstein model using two methodologies: (i) a static mean-field approach, and (ii) the cluster perturbation theory -- the latter provides the spectral functions of an infinite system with short-range correlations, i.e.~beyond a Hartree-Fock approximation.
Given this, our main findings are summarized in the phase diagram of Fig.\,\ref{phasediagram}, which presents the CPT and MFT results, from which one notices that the RSOC is harmful to the CDW phase.
This deleterious feature of the RSOC could be understood from the diminishing DOS at the half-filling when $t_R$ increases, from a weak coupling point-of-view; or by a large energetic price to pay to avoid spin-flip processes (within the CDW phase), from a strong coupling point-of-view.
Furthermore, as shown in Figs.\,\ref{fig:DOS_plaquete_lamb1.2} and \ref{fig:CDW_Gap}, the transition from the CDW phase to a \textit{correlated} Rashba metal seems to be continuous, with the transition line being pushed to larger values of $t_R$ when $\omega_{0} \to 0$.
Even though the critical phenomena properties (such as critical exponents) are beyond the scope of this work,
our phase diagram $g \times t_{\rm R}$ provides accurate critical points, whose fundamental properties may shed light into the growing class of Rashba materials \cite{CastroNeto-Led-Chalcogenide-PhysRevB.96.161401, CastroNeto-Led-Chalcogenide-PhysRevB.97.235312, 2D-SOC-Review-Chen2021}. 

\section*{ACKNOWLEDGMENTS}
The authors are grateful to R.R.~dos Santos for enlightening discussions and suggestions.
Financial support from Fundação Carlos Chagas Filho de Amparo à Pesquisa do Estado do Rio de
Janeiro, grant number E-26/200.258/2023 - SEI-260003/000623/2023 (N.C.C.); and from CNPq grant number 313065/2021-7 (N.C.C.) is gratefully acknowledged.
We also acknowledge support from INCT-IQ. T.P.C. acknowledge support from CAPES (Brazil). S.A.S.Jr. acknowledges support from CNPq (Brazil) and Fundação Carlos Chagas Filho de Amparo à Pesquisa do Estado do Rio de Janeiro process SEI-260003/006232/2023. R.A.F. acknowledges support from Fundação Carlos Chagas Filho de Amparo à Pesquisa do Estado do Rio de Janeiro and CNPq (Brazil).

\bibliography{ref}
\end{document}